\documentclass[12pt,showpacs,amsmath,amssymb]{revtex4}%
\DeclareRobustCommand{\baselinestretch{1}}

\usepackage{color}
\usepackage{epic}
\usepackage{eepic}
\usepackage{graphicx}

\begin{document}

\title{ON MODELING OF STATISTICAL PROPERTIES OF CLASSICAL  3$D$ SPIN GLASSES }

\author{A.~S.~Gevorkyan$^{1,2}$, H.~G.~ Abajyan$^{1}$ and  E. A. Ayryan$^{2}$}
\email[]{g_ashot@sci.am}\quad \email[]{habajyan@ipia.sci.am}\quad
\email[]{ayrjan@jinr.ru}
 \affiliation{$^1$Institute for Informatics and Automation Problems, NAS of
Armenia, 0014, 1 P. Sevak, Yerevan, Armenia\\ $^2$Joint Institute of
Nuclear Research,  141980 Dubna, Moscow reg., Russia}

\begin{abstract}\textbf{ABSTRACT}\\
We study statistical properties of  3$D$ classical spin glass layer
of certain width and infinite length. The 3$D$ spin glass is
represented as an ensemble of disordered 1$D$ spatial spin-chains
(SSC) where interactions are random between spin-chains (nonideal
ensemble of 1$D$ SSCs). It is proved that at the limit of Birkhoff's
ergodic hypothesis performance 3$D$ spin glasses can be generated by
Hamiltonian of disordered 1$D$ SSC with random environment.
Disordered 1$D$ SSC is defined on a regular lattice where one
randomly oriented spin is put on each node of lattice. Also it is
supposed that each spin randomly interacts with six
nearest-neighboring spins (two spins on lattice and four in the
environment). The recurrent transcendental equations are obtained on
the  nodes of spin-chain lattice. These equations combined with the
Silvester conditions allow step by step construct spin-chain  in the
ground state of energy  where all spins are in minimal energy of
classical Hamiltonian. On the basis of these equations  an original
high-performance parallel algorithm is developed for 3$D$ spin
glasses simulation. Distributions of different parameters of
unperturbed spin glass are calculated. In particular, it is
analytically proved and by numerical calculations shown that the
distribution of spin-spin interaction constant in Heisenberg
nearest-neighboring Hamiltonian model as opposed to widely used
Gauss-Edwards-Anderson distribution satisfies L\'{e}vy alpha-stable
distribution law which does not have variance.  A new formula is
proposed for construction of partition function in kind of
one-dimensional integral on energy distribution of 1$D$ SSCs.
\end{abstract}


 \maketitle

\textbf{keywords}: {Spin-glass Hamiltonian, random network, Birkhoff
ergodic hypothesis, statistic distributions,  parallel algorithm,
numerical simulation.}

\vspace{3.8 cm}
\section{Introduction}
 The wide class of phenomena and structures in
physics, chemistry, material science, biology, nanoscience,
evolution, organization dynamics, hard-optimization, environmental
and social structures, human logic systems, financial mathematics,
etc. are mathematically well described in the framework of spin
glass models \cite{Bind,Mezard,Young,Ancona,Bov,Fisch,Tu,Chary,Baake}.\\
The considered mean-field models of spin glasses as a rule are
divided into two types. The first consists of the true random-bond
models where the coupling  between interacting spins are taken to be
independent random variables  \cite{Sherington,Derrida,Parisi}.
 The solution of these models is obtained by $n$-replica trick
\cite{Sherington,Parisi} and requires invention of sophisticated
schemes of  replica-symmetry breaking \cite{Parisi,Bray}. In the
models of second type the bond-randomness is expressed in terms of
some underlining hidden site-randomness and is thus of a superficial
nature. It has been pointed out in the works
\cite{Fernand,Benamir,Grensing}, however, this feature retains an
important physical aspect of true spin glasses, viz. they are random
with respect to the positions of magnetic impurities.\\
Note that all mentioned investigations as a rule conduct at
equilibrium's conditions of medium. This fact plays a key role both
in analytical and numerical simulation by Monte Carlo
method.\\
Recently, as authors have shown \cite{gev} some type of dielectrics
can be studied by model of quantum 3$D$ spin glass. In particular,
it was proved that the initial 3$D$ quantum problem on scales of
space-time periods of an external fields can be reduced to two
conditionally separable 1$D$ problems where one of them describes an
ensemble of disordered 1$D$ spatial spin-chains between which are
random interactions (further  will be called \emph{\textbf{nonideal
ensemble}}).\\
In this paper we discuss in detail statistical properties of
classical 3$D$ spin glass with suggestion that interactions between
spins have short-range character. We  prove that nonideal ensemble
of 1$D$ SSCs exactly describes the statistical properties of
classical 3$D$ spin glasses in the limit of Birkhoff's ergodic
hypothesis performance. In the work a new high performance algorithm
for simulation of this traditionally difficult calculated problem is
developed.
\begin{figure}\center
\includegraphics[height=65mm,width=125mm]{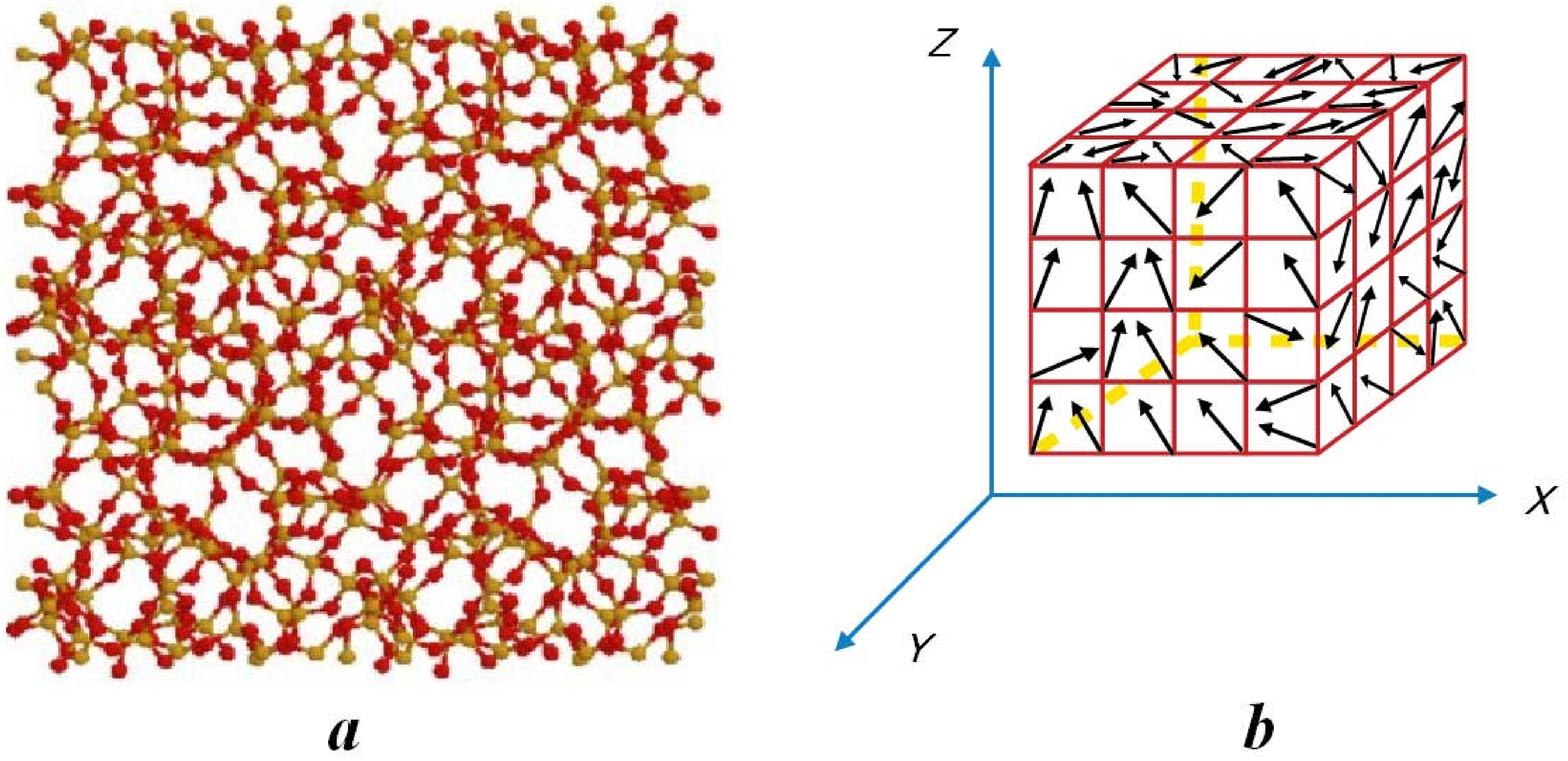}
 \caption{The structure of amorphous silicon dioxide $SiO_2$ is
described by $3D$ random network with covalent bonds. Every silicon
vertex (gold sphere) has 4 edges and every oxygen vertex (red
sphere) has 2 edges.}
 \label{fig1}
\end{figure}

In section II the classical  spin glass problem on 3$D$ lattice is
formulated. Equations for stationary points and corresponding
Silvester  conditions  are obtained for definition of  energy
minimum on lattice nodes (local minimum of energy). The formula for
computation of different parameters distributions of spin glass is
defined.

In section III a theorem on reduction of 3$D$ spin glass problem to
the problem of nonideal ensemble of 1$D$ SSCs  is proved.

In section IV numerical experiments are adduced for unperturbed 1$D$
SSCs ensemble with spin-chain's length $10^3d_0$. In particular,
distributions of energy,  polarization and spin-spin interaction
constants of nonideal ensemble are investigated in detail.

In section V partition function is investigated  in detail in the
configuration integral's representation. A new representation  is
suggested for partition function in kind of one dimensional integral
on energy distribution of nonideal ensemble.

In section VI the obtained theoretical and computational results are
analyzed. It is very important to note that it has been proved that
in the framework of the developed method it is always possible to
exactly compute the ground state energy of 3$D$ spin glasses.

\section{Formulation of Problem}
The objects of our investigation are  solid-state dielectrics, type
of $SiO_{2}$ glass (amorphous silicon dioxide). According to the
numerical $ab$ $initio$ simulations \cite{Tu}, the structure of this
type compound can be well described by $3D$ random network (Fig.
1a). The red and brown lattice points on this figure correspond to
different atoms, while the links between them correspond to covalent
bounds.  As a result of charges redistribution in outer electronic
shells,  atoms of $Si$ acquire the positive charge and atoms of $O$
correspondingly  the negative charge. Thus, we can consider
compounds of this type as a disordered $3D$ system of similar rigid
dipoles (hereinafter termed as a system of $3D$ disordered spins,
Fig. 1b). Let us remind that under the similar rigid dipoles are
meant the dipoles for which the absolute values are equal
($|{\bf{p}}_i|=|{\bf{p}}_j|=p^{\,0}$, where ${\bf{p}}_i$ and
${\bf{p}}_j$ are two arbitrary dipoles), and they don't vary under
the influence of an external field.

The Hamiltonian of 3$D$ classical spin glass system  reads:
 $$
 H(\{\textbf{r}\})=-\sum_{<i\,j>}
J_{i\,j}\textbf{\emph{S}}_{i}\textbf{\emph{S}}_{j},\qquad
\{\textbf{r}\}\equiv \textbf{r}_1,\textbf{r}_2,....
$$
where indices $i$ and $j$  run over all nodes of 3$D$ lattice,
 $\textbf{r}_i$ correspondingly denotes the  coordinates of $i$-th spin (see Fig. 1b).
For further investigation we will consider a spin glass layer of
certain width $L_x$ and infinite length (see Fig. 2).  We will
consider 3$D$ compound in the framework of nearest-neighboring
 Hamiltonian model. Let us note that even for this relatively
simplest model numerical simulations of spin glasses  are extremely
hard to solve NP  problems.

At first we will consider an auxiliary Heisenberg Hamiltonian of the
form:
\begin{eqnarray}
H_0(\{\textbf{r}\};\,
N_x)=H_0^{(1)}(\{\textbf{r}\};\,N_x)+H_0^{(2)}(\{\textbf{r}\};\,N_x),
\label{01}
\end{eqnarray}
where the first term $H_0^{(1)}(\{\textbf{r}\};\,N_x)$:
 $$
 H_0^{(1)}(\{\textbf{r}\};\,N_x)=-\sum_{i=0}^{N_x-1}
J_{i\,i+1}\textbf{\emph{S}}_{i}\textbf{\emph{S}}_{i+1},
$$
describes the disordered 1$D$ \emph{spatial spins chain} (SSC) while
the second term $H_0^{(2)}(\{\textbf{r}\};\,N_x)$:
$$ H_0^{(2)}(\{\textbf{r}\};\,N_x)=-\sum_{i=0}^{N_x-1}\sum_{\sigma=1}^{4}
J_{i\,i_\sigma}\textbf{\emph{S}}_{i}\textbf{\emph{S}}_{i_\sigma},
$$
correspondingly describes the random surroundings of 1$D$ SSC (see
Fig. 2).  In (\ref{01}) $J_{i\,i+1}$ and $J_{i\,i_\sigma}$ are
correspondingly random interaction constants between arbitrary $i$
and $i+1$ spins and  between $i$ and $i_\sigma$ spins,
$\textbf{\emph{S}}_{i},\,\textbf{\emph{S}}_{i+1}$ and
$\textbf{\emph{S}}_{i_\sigma}$ are spins (vectors) of unit length,
which are randomly orientated in O(3) space. From the general
reasons it follows that with the help of (\ref{01}) Hamiltonian and
 by way of successive constructing we can restore the Hamiltonian of 3$D$ problem.
 Recall that the meaning of the construction is as follows. On
the first step the central spin-chain on the $x$-axis with its
surroundings from four random spin-chains is considered (see Fig.
2). On the second step as central spin-chains are considered
corresponding spin-chains from the random surroundings each of which
are surrounded by new four neighboring spin-chains. Thus, repeating
this cycle periodically we can construct the Hamiltonian of 3$D$
problem. This idea will be rigorously proved below.
\begin{figure}\center
\includegraphics[height=80mm,width=120mm]{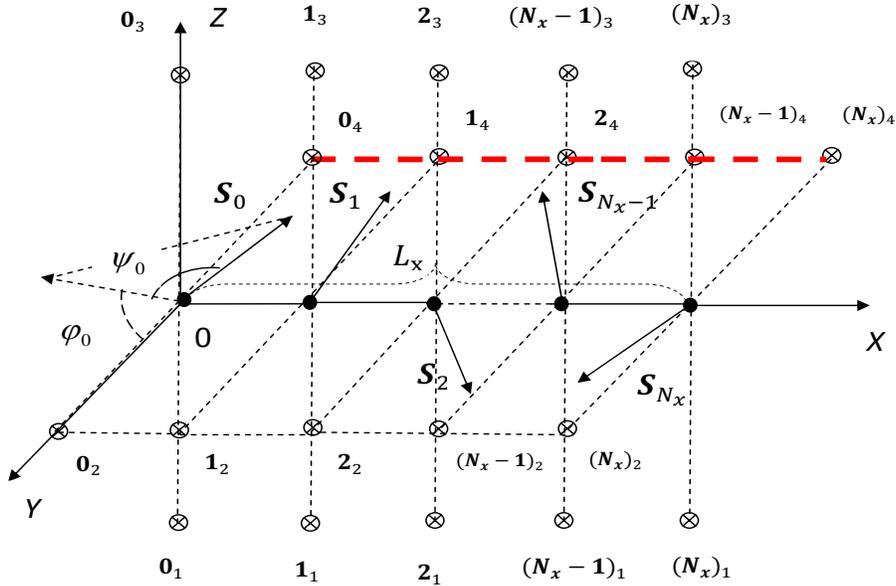}
\caption{1$D$ SSC with the random environment. Recall that each
spin-chain is surrounded by four spin-chains which are randomly
interacted with it. By symbols $\otimes$ are designated spins from
the random environment (four spin-chains of surrounding).}
\end{figure}

For further investigation of spin glass problem it is useful to
write Hamiltonian (\ref{01}) in spherical coordinates system:
\begin{eqnarray}
H_0(\{\textbf{r}\};\,N_x)=-\sum_{i=0}^{N_x-1}\biggl\{
J_{i\,i+1}\Bigl[\cos\psi_i\cos\psi_{i+1}\cos(\varphi_i-\varphi_{i+1})
+\sin\psi_i\sin\psi_{1+1}\Bigr]\nonumber\\
\qquad\qquad\,\,+\,\,\sum_{\sigma=1}^{4}
J_{i\,i_\sigma}\Bigl[\cos\psi_i\cos\psi_{i_\sigma}
 \cos(\varphi_i-\varphi_{i_\sigma})+\sin\psi_i\sin\psi_{i_\sigma}\Bigr]\biggr\}.
 \label{02}
\end{eqnarray}
Now the main problem is to find the angular configurations and
spin-spin interaction constants which can make the Hamiltonian
minimal on each node of lattice.

 Let us consider the equations of stationary point:
\begin{equation}
\frac{\partial{H_0}}{\,\,\partial\psi_i}=0,\qquad\qquad
\frac{\partial{H_0}}{\,\,\partial\varphi_i}=0, \label{03}
\end{equation}
where ${\Theta }_{i}=(\psi _{i},\varphi _{i})$ defines the
orientation of $i$-th spin ($\psi _{i},\varphi _{i}$ are
correspondingly the polar and the azimuthal angles).  In addition,
$\mathbf{\Theta }=({ \Theta_{1}},{\Theta _{2}}....{\Theta
_{N_{x}}})$ describes the angular configuration of spin-chain
consisting of $N_x$ spins.

Substituting (\ref{02}) into (\ref{03}) we can find the following
recurrent equations:
$$
 J_{i-1\,i}\,\Bigl[-\sin\psi_i\cos\psi_{i-1}\cos(\varphi_i-\varphi_{i-1})
+\cos\psi_i\sin\psi_{i-1}\Bigr]+\nonumber\\
$$
$$
J_{i\,i+1}\,\Bigl[-\sin\psi_i\cos\psi_{i+1}\cos(\varphi_i-\varphi_{i+1})
+\cos\psi_i\sin\psi_{i+1}\Bigr]+
$$
$$
\sum_{\sigma=1}^{4}J_{i\,i_\sigma}\Bigl[-\sin\psi_i\cos\psi_{i_\sigma}\cos(\varphi_i-\varphi_{i_\sigma})
+\cos\psi_i\sin\psi_{i_\sigma}\Bigr]=0,
 \vspace{0.150 cm}
$$
\begin{eqnarray}
\quad\Bigl\{\,\Bigl[\,J_{i-1\,i}\,\cos\psi_{i-1}\sin(\varphi_i-\varphi_{i-1})+
J_{i\,i+1}\,\cos\psi_{i+1}\,\times
\nonumber\\
\quad\sin(\varphi_i-\varphi_{i+1})\Bigr]+\sum_{\sigma=1}^{4}J_{i\,i_\sigma}\cos\psi_{i_\sigma}
\sin(\varphi_i-\varphi_{i_\sigma})\Bigr\}\cos\psi_i=0. \label{04}
\end{eqnarray}

 In order to satisfy the conditions of local minimum (Silvester conditions) for $H_0$, it is
necessary that the following inequalities are carried out:
\begin{equation}
A_{\psi_i\psi_i}({\Theta^0_i})>0,\quad
A_{\psi_i\psi_i}({\Theta^0_i})A_{\varphi_i\varphi_i}({\Theta^0_i})
-A_{\psi_i\varphi_i}^2({\Theta^0_i})>0,
 \label{05}
\end{equation}
where $A_{\alpha_i\alpha_i} ={\partial^2{H_0}}/{\partial\alpha_i^2}$
and $
A_{\alpha_i\beta_i}=A_{\alpha_i\beta_i}={\partial^2{H_0}}/{\partial\alpha_i\partial\beta_i}$,
in addition:
$$
 A_{\psi_i\psi_i}({\Theta^0_i})\,
=J_{i-1\,i}\Bigl\{\cos\psi_i^0\cos\psi_{i-1}\cos(\varphi_i^0-\varphi_{i-1})
+\sin\psi_i^0\sin\psi_{i-1}\Bigr\}
+J_{i\,i+1}\Bigl\{\cos\psi_i^0\cos\psi_{i+1}\times
$$
$$
\cos(\varphi_i^0-\varphi_{i+1}) +\sin\psi_i^0\sin\psi_{i+1}\Bigr\}
+\sum_{\sigma=1}^{4}J_{i\,i_\sigma}\Bigl\{\cos\psi_i^0\cos\psi_{i_\sigma}
\cos(\varphi_i^0-\varphi_{i_\sigma})+\sin\psi_i^0\sin\psi_{i_\sigma}\,\Bigr\},
$$
$$
A_{\varphi_i\varphi_i}({\Theta^0_i})=
\Bigl\{J_{i-1\,i}\cos\psi_{i-1}\cos(\varphi_i^0-\varphi_{i-1})
+J_{i\,i+1}\cos\psi_{i+1}\cos(\varphi_i^0-\varphi_{i+1})\,
+\,\sum_{\sigma=1}^{4}J_{i\,i_\sigma}\times
$$
$$
\cos\psi_{i_\sigma}\cos(\varphi_i^0-\varphi_{i_\sigma})\Bigr\}\cos\psi_i^0,\qquad
A_{\psi_i\varphi_i}({\Theta^0_i})=0.
$$
Recall that $\Theta^0_i=(\psi_i^0,\varphi_i^0)$ designates the
angular configuration of the spin in case when the condition of
local minimum for $H_0$ is satisfied.

Thus, it is obvious that the classical 3$D$ spin glass system (see
Fig. 1b) can be considered as an nonideal ensemble of 1$D$ SSCs (see
Fig. 2) and there are random interactions between spin-chains.

Now we can construct distribution functions of different parameters
of 1$D$ SSCs nonideal ensemble. To this effect it is useful to
divide the nondimensional energy axis
$\varepsilon=\epsilon/\delta\epsilon$ into regions $0>\varepsilon
_{0}>...>\varepsilon_{n}$, where $n>>1$ and $\epsilon$ is the real
energy axis. The number of stable 1$D$ SSC configurations with
length $L_{x}$ in the range of energy $ [\varepsilon -\delta
\varepsilon,\varepsilon +\delta \varepsilon ]$ will be denoted by
$M_{L_{x}}(\varepsilon )$ while the number of all stable 1$D$ SSC
configurations - correspondingly by symbol $M_{L_{x}}^{full}=
\sum_{j=1}^{n}M_{L_{x}}(\varepsilon_{j})$. Accordingly, the energy
distribution function  can be defined by the expression:
\begin{equation}
F_{L_x}(\varepsilon;d_0(T))=M_{L_x}(\varepsilon)/M_{L_x}^{full},
\label{06}
\end{equation}
 where distribution function is normalized to unit:
$$
 \lim_{n\to\infty}\sum^n_{j=1}
F_{L_x}(\varepsilon_j;d_0(T))\delta \varepsilon_j=
\int^{\,0}_{-\infty}F_{L_x}(\varepsilon;d_0(T))d\varepsilon=1.
$$
By similar way we  can construct also distribution functions for
polarizations, spin-spin interaction constant, etc.

\section{Reduction of 3$D$ spin glass problem to 1$D$ SSCs ensemble problem }

Modeling of 3$D$ spin glasses is a typical NP hard problem. This
type of problems are hard-to-solve even on modern supercomputers if
the number of spins in the system are  more or less significant. In
connection with  told, the significance of new mathematical
approaches development is obvious and on the basis of which an
effective parallel algorithms for numerical simulation of spin
glasses can be elaborated.

{\bf Theorem}: \emph{The  classical} 3$D$ \emph{spin glass problem
at the limit  of isotropy and homogeneity  (ergodicity) of
superspins distribution (sum of spins in chain) in} 3$D$
\emph{configuration space is equivalent to the problem of
disordered} 1$D$ \emph{SSCs ensemble. }

It is obvious that the theorem will be  proved if we can prove that
 in case when the distribution of superspins in 3$D$
configuration space is homogeneous and isotropic, the following two
propositions take place:
\\
\textbf{\emph{a}}) In any random environment which consists of four
arbitrary  spin-chains it is always possible to find at least one
physically admissible solution for spin-chain  (the direct problem), and\\
\textbf{\emph{b}}) It is possible to surround an arbitrary
spin-chain from the given environment with such environment which
can make it physically admissible spin-chain solution (the reverse
problem).

The direct Problem.\\ Using the following notation:
\begin{equation}
\xi_{i+1} =\cos\psi_{i+1},\qquad
\eta_{i+1}=\sin(\varphi_i-\varphi_{i+1}),
 \label{07}
\end{equation}
equations system (\ref{06}) can be transformed to the following
form:
\begin{eqnarray}
C_1+J_{i\, i+1}\bigl[\sqrt{1-\xi^2_{i+1}}-\tan\psi_i\,\xi_{i+1}
\sqrt{1-\eta^2_{i+1}} \bigr]=0,\nonumber\\
C_2+J_{i\, i+1}\,\xi_{i+1}\,\eta_{i+1}=0,
 \label{08}
\end{eqnarray}
where parameters $C_1$ and $C_2$ are defined by expressions:
$$
C_1=J_{i-1\,i}\bigl[\sin\psi_{i-1}-\tan\psi_i\cos\psi_{i-1}\cos(\varphi_i-\varphi_{i-1})]
+\sum_{\sigma=1}^{4}J_{i\,i_\sigma}\times \vspace{-0.5 cm}
$$
$$
\Bigl[\sin\psi_{i_\sigma}-\tan\psi_i\cos\psi_{i_\sigma}\cos(\varphi_i-\varphi_{i_\sigma})
\Bigr],\qquad\qquad\quad\,\,
$$
$$
 C_2=J_{i-1\,i}\cos\psi_{i-1}
\sin(\varphi_i-\varphi_{i-1})+
 \sum_{\sigma=1}^{4}J_{i\,i_\sigma}\cos\psi_{i_\sigma}
\sin(\varphi_i-\varphi_{i_\sigma}).
$$
From the system (\ref{08}) we can find the equation for the unknown
variable $\eta_{i+1}$:
\begin{equation}
C_1\eta_{i+1}+C_2\sqrt{1-\eta^2_{i+1}}\tan\psi_i+\sqrt{J_{i\,
i+1}^2\eta^2_{i+1}-C_2^2}=0.
 \label{09}
\end{equation}
 We have transformed the equation (\ref{09}) to the equation of
fourth order which is exactly solved further:
\begin{equation}
\xi^2_{i+1} =\frac{C_2^2}{J_{i\,i+1}^2\eta^2_{i+1}}, \qquad
\eta^2_{i+1} = \frac{A}{B},
 \label{10}
\end{equation}
where
$$
A=C_2^2\Bigl\{J_{i\,i+1}^2\cos^2\psi_i+
C_3+2C_1^2\sin^2\psi_i\Bigl[1\, \pm\, C_1^{-1}\sqrt{J_{i\,i+1}^2-
C_1^2-C_2^2}\,\cot\psi_i\Bigr]\Bigr\}.\qquad\quad$$
$$
 C_3=-C_1^2+C_2^2\,\sin^2\psi_i,\quad
B=J_{i\,i+1}^4\cos^4\psi_i+
2C_3J_{i\,i+1}^2\cos^2\psi_i+(C_1^2+C_2^2\sin^2\psi_i)^2,
$$
 Note that from the condition of nonnegativity of the value under the root
we can find the following nonequality:
\begin{equation}
\qquad \qquad\qquad J_{i\,i+1}^2 \geq C^2_1 + C^2_2.
 \label{11}
 \end{equation}
In consideration of (\ref{07}), we can write following conditions:
$$
0\leq \xi^2_{i+1}\leq 1, \qquad 0\leq \eta^2_{i+1}\leq 1.
$$
As it follows from  equations (\ref{10}), if the solutions in
previous two nodes $(i-1)$ and $i$ are known, then the solutions
$(\psi_{i+1},\varphi_{i+1})$ in the node $(i+1)$ can be defined only
by  constant  $J_{i\,i+1}$. In this connection a natural question
arises - are there solutions for spin-chain in arbitrarily given
environment?

Let us consider Silvester conditions (\ref{05}) which can be written
in the form of the following inequalities:
\begin{eqnarray}
J_{i\,i+1}\cos\psi_i^0\cos\psi_{i+1}\cos(\varphi_i^0-\varphi_{i+1})
>
-a_1-\sin\psi_i^0\sin\psi_{i+1} , \nonumber\\
J_{i\,i+1}\cos\psi_{i+1}\cos(\varphi_i^0-\varphi_{i+1})\cos\psi_i^0
> -a_2,
 \label{12}
\end{eqnarray}
where constants $a_1$ and  $a_2$ are defined by expressions:
$$
 a_1 =J_{i-1\,i}\Bigl[\cos\psi_i^0\cos\psi_{i-1}\cos(\varphi_i^0-\varphi_{i-1})
+\sin\psi_i^0\sin\psi_{i-1}\Bigr]+\sum_{\sigma=1}^{4}J_{i\,i_\sigma}\Bigl[\cos\psi_i^0\times
$$
$$
\cos\psi_{i_\sigma}\cos(\varphi_i^0-\varphi_{i_\sigma})\,+\,\sin\psi_i^0\sin\psi_{i_\sigma}\,\Bigr],
$$
$$
a_2\,=\,\Bigl\{J_{i-1\,i}\cos\psi_{i-1}\cos(\varphi_i^0-\varphi_{i-1})
\,+ \,\sum_{\sigma=1}^{4}J_{i\,i_\sigma}
\cos\psi_{i_\sigma}\cos(\varphi_i^0-\varphi_{i_\sigma})\Bigr\}\cos\psi_i^0.\,\,\,
$$
So, the problem leads to the answer of the following question - are
inequality (\ref{11}) and (\ref{12}) compatible or not. Taking into
account solutions (\ref{10}) it is easy to prove that conditions
(\ref{12}) are automatically compatible at large absolute values of
$J_{i\,i+1}$. On the other hand, there is no any contradiction with
condition (\ref{11}).
\begin{figure}\center
\includegraphics[height=70mm,width=70mm]{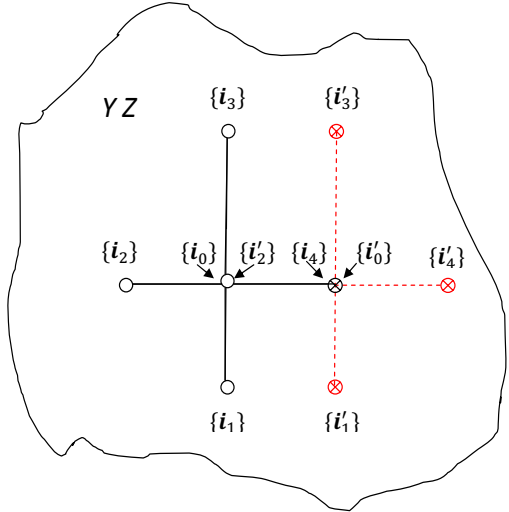}
\caption{The projection of spin-chains ensemble on the $(Y,Z)$
plane. Spin-chains are designated by symbols $\bigcirc$ and
$\bigotimes$ which correspondingly form the old and new
environments.}
\end{figure}
 Thus the direct problem or the proposition \textbf{\emph{a}}) is
 proved.\\
Now our aim is to prove the reverse problem or the proposition
\textbf{\emph{b}}) which consists of the following. We choose a
spin-chain from the environment (see Fig. 2), for example
$\{i_4\}\equiv(0_4,1_4,...,{N_x}_4)$. In this spin-chain all angular
configurations of spins $(\Theta_0^{(4)},...\Theta_{N_x}^{(4)})$ are
known  but the constants that define  spin-spin interactions in
spin-chain and interactions between spin-chain and its environment
still are not defined. We will prove that it is always possible to
surround each spin-chain by such environment that the selected
spin-chain will be the correct solution from the main physical laws
point of view (see conditions (\ref{04})-(\ref{05})). In the
considered case $\{i_4\}\equiv \{i_0^{'}\}$ spin-chain is surrounded
by four neighbors, one of which $\{i_0\}\equiv \{i_2^{'}\}$ is fully
determined while three spin-chains $\{i_1^{'}\},\{i_3^{'}\}$ and
$\{i_4^{'}\}$ should be still specified (see Fig. 3). Recall that
the mark "$^{\,'\,}$" designates a new environment with three
spin-chains. However, for simplicity we will omit or more clearly
make change them in the subsequent calculations
$\bigl(\{i_0^{'}\},\{i_1^{'}\},\{i_2^{'}\},\{i_3^{'}\},\{i_4^{'}\}\bigr)\to
\bigl(\{i_0\},\{i_1\}\{i_2\},\{i_3\},\{i_4\}\bigr)$. The proof of
the proposition  should be conducted as follows. We will suppose
that the constants of spin-spin interactions in considered chain and
corresponding  parameters of two spin-chains of environment are
known.  We will show that by special choosing of parameters of the
third spin-chain $\{i_3\}$ it is possible to ensure the condition of
local minimum energy is satisfied in the considered spin-chain.

 Let us define the following denotations for constants:
\begin{eqnarray}
c_1=J_{i-1\,i}[-\sin\psi_i\cos\psi_{i-1}\cos(\varphi_i-\varphi_{i-1})+\cos\psi_i\sin\psi_{i-1}]+\nonumber\\
\qquad
\,\,J_{i\,i_\sigma}\,[-\sin\psi_i\cos\psi_{i_\sigma}\cos(\varphi_i-\varphi_{i_\sigma})+
\cos\psi_i\sin\psi_{i_\sigma}],
\nonumber\\
c_2=-\sin\psi_i\cos\psi_{i+1}\cos(\varphi_i-\varphi_{i+1})+\cos\psi_i\sin\psi_{i+1},\nonumber\\
c_3=J_{i-1\,i}\cos\psi_{i-1}\sin(\varphi_i-\varphi_{i-1})
+J_{i\,i_\sigma}\cos\psi_{i_\sigma}\sin(\varphi_i-\varphi_{i_\sigma}),\nonumber\\
c_4=\cos\psi_{i+1}\sin(\varphi_i-\varphi_{i+1}),\qquad \sigma=4.
\label{13}
\end{eqnarray}
Using (\ref{13}) we can transform equations (\ref{04}) to the
following form:
$$
c_1+c_2J_{i\,i+1}+\sum_{\sigma =
1}^{3}{J_{i\,i_\sigma}}[-\sin\psi_i\cos\psi_{i_\sigma}\cos(\varphi_i-
\varphi_{i_\sigma})+\cos\psi_i\sin\psi_{i_\sigma}]=0,
$$
$$
c_3+c_4J_{i\,i+1}+ \sum_{\sigma =
1}^{3}{J_{i\,i_\sigma}}\cos\psi_{i_\sigma}\sin(\varphi_i-\varphi_{i_\sigma})=0,
\qquad\qquad\qquad\qquad\qquad
$$
which are equivalent to the following relations:
\begin{eqnarray}
J_{i\,i+1}= -\frac{c_1}{c_2}-\frac{1}{c_2}\sum_{\sigma =
1}^{3}{J_{i\,i_\sigma}}[-\sin\psi_i\cos\psi_{i_\sigma}\cos(\varphi_i-\varphi_{i_\sigma})+
\cos\psi_i\sin\psi_{i_\sigma}],
\nonumber\\
J_{i\,i+1}= -\frac{c_3}{c_4}-\frac{1}{c_4}\sum_{\sigma =
1}^{3}{J_{i\,i_\sigma}}\cos\psi_{i_\sigma}\sin(\varphi_i-\varphi_{i_\sigma}).
\label{14}
\end{eqnarray}
After excluding $J_{i\,i+1}$ from (\ref{14}) we find the following
equation:
\begin{eqnarray}
\sum_{\sigma=1}^{3}\biggl\{{\frac{J_{i\,i_\sigma}}{c_2}}[-\sin\psi_i\cos\psi_{i_\sigma}
\cos(\varphi_i-\varphi_{i_\sigma})+\cos\psi_i\sin\psi_{i_\sigma}]
-\nonumber\\
\qquad{\frac{J_{i\,i_\sigma}}{c_4}}\cos\psi_{i_\sigma}\sin(\varphi_i-
\varphi_{i_\sigma})\biggr\}-c_5=0,\qquad c_5=\frac{c_1}{c_2}
-\frac{c_3}{c_4}.  \label{15}
\end{eqnarray}
Having made the following designation:
$$
D=\sum_{\sigma
=1}^{2}\biggl\{\frac{J_{i\,i_\sigma}}{c_2}[-\sin\psi_i\cos\psi_{i_\sigma}
\cos(\varphi_i-\varphi_{i_\sigma}) +\cos\psi_i\sin\psi_{i_\sigma}]$$
$$
-\frac{J_{i\,i_\sigma}}{c_4}\cos\psi_{i_\sigma}\sin(\varphi_i-
\varphi_{i_\sigma})\biggr\}-c_5,\qquad\qquad
$$
we can transform equation (\ref{15}) to the following form:
\begin{eqnarray}
D+{\frac{J_{i\,i_3}}{c_2}}[-\sin\psi_i\cos\psi_{i_3}\cos(\varphi_i-\varphi_{i_3})
+\cos\psi_i\sin\psi_{i_3}]
\nonumber\\
\,\quad-\,{\frac{J_{i\,i_3}}{c_4}}\cos\psi_{i_3}\sin(\varphi_i-\varphi_{i_3})=0.
\label{16}
\end{eqnarray}
Now substituting:
\begin{equation}
 x =\cos\psi_{i_3},
 \label{17}
\end{equation}
in (\ref{16}) we find the equation:
\begin{eqnarray}
D+{\frac{J_{i\,i_3}}{c_2}}[-x\sin\psi_i\cos(\varphi_i-\varphi_{i_3})+\sqrt{1-{x}^2}\,\cos\psi_i]
\nonumber\\
-x{\frac{J_{i\,i_3}}{c_4}}\sin(\varphi_i-\varphi_{i_3})=0,
\label{18}
 \end{eqnarray}
From (\ref{18}) the following square equation can be found :
\begin{eqnarray}
K_0x^2+2K_1x+K_2=0,
 \label{19}
\end{eqnarray}
where the following designations are made:
$$
K_0=
\cos^2\psi_i+\biggl(\sin\psi_i\cos(\varphi_i-\varphi_{i_3})+\frac{c_2}{c_4}\sin(\varphi_i-\varphi_{i_3})
\biggr)^2,
$$
$$
K_1=-\frac{Dc_2}{J_{i\,i_3}}\biggl(\sin\psi_i\cos(\varphi_i-\varphi_{i_3})+
\frac{c_2}{c_4}\sin(\varphi_i-\varphi_{i_3})\biggr),\quad K_2=
\biggl(\frac{Dc_2}{J_{i\,i_3}}\biggr)^2-\cos^{2}\psi_i.
$$
Discriminant of the square equation (\ref{19}) has the form:
\begin{eqnarray}
D_x=\biggl(\sin\psi_i\cos(\varphi_i-\varphi_{i_3})+\frac{c_2}{c_4}\sin(\varphi_i-\varphi_{i_3})
\biggr)^2 \cos^2\psi_i \nonumber\\
\quad\qquad+\biggl\{\cos^2\psi_i-\biggl(\frac{Dc_2}{J_{i\,i_3}}\biggr)^2\biggr\}\cos^2\psi_i\geq
0,
 \label{20}
\end{eqnarray}
which on  some set of $J_{i\,i_3}$ can be positive, i.e. $i$-th spin
in spin-chain $\{i_4\}$ will satisfy the local minimum conditions.

 Let us  define:
\begin{equation}
 y =\cos(\varphi_i-\varphi_{i_3}),
 \label{21}
\end{equation}
Substituting (\ref{21}) in  (\ref{16}) we will find that:
$$
D+{\frac{J_{i\,i_3}}{c_2}}[-y\,\sin\psi_i\cos\psi_{i_3}+\cos\psi_i\sin\psi_{i_3}]
-{\frac{J_{i\,i_3}}{c_4}}\cos\psi_{i_3}\sqrt{1-y^2}=0,
$$
After squaring we will have the following equation:
\begin{eqnarray}
M_0y^2+2M_1y+M_2=0, \label{22}
\end{eqnarray}
where the following designations are made:
$$
M_0=
\biggl(\biggl(\frac{c_2}{c_4}\biggr)^2+\sin^2\psi_i\biggr)\cos^2\psi_{i_3},\quad
M_1=-\sin\psi_i\cos\psi_{i_3}\biggl(\cos\psi_i\sin\psi_{i_3}+\frac{Dc_2}{J_{i\,i_3}}\biggr),
$$
$$
M_2=\biggl(\cos\psi_i\sin\psi_i+ \frac{Dc_2}{J_{i\,i_3}}\biggr)^2
-\biggl(\frac{c_2}{c_4}\biggr)^2 \cos^2\psi_{i_3}.
$$
 The discriminant of the square equation (\ref{22}) has the form:
\begin{eqnarray}
D_y=\biggl(\frac{c_2}{c_4}\biggr)^2\cos^{2}\psi_i+\sin^{2}\psi_i\cos^{2}\psi_{i_3}
-
\biggl(\frac{Dc_2}{J_{i\,i_3}}+\cos\psi_i\sin\psi_{i_3}\biggr)^{2}\geq
0. \label{23}
\end{eqnarray}
Obviously there are some set of constants $J_{i\,i_3}$ on which
$D_y\geq 0$. However, it is  more important to find the region of
the interaction constant $J_{i\,i_3}$ values for which both
determinants $D_x$ and $D_y$ are positive.

In particular as the analysis of the following condition shows:
\begin{equation}
-\biggl|\frac{Dc_2}{\cos\psi_i}\biggr| \geq J_{i\,i_3} \geq
\biggl|\frac{Dc_2}{\cos\psi_i}\biggr|, \label{24}
\end{equation}
 discriminant $D_x$  is always nonnegative. From the other side:
\begin{equation}
\sin\psi_{i_3} \cong -\frac{D{c_2}}{{J_{i\,i_3}}{\cos\psi_i}},
\label{25}
\end{equation}
which will assure that $D_y$ discriminant is always nonnegative. A
simple analysis of conditions (\ref{24}) and (\ref{25}) shows  that
they  are compatible. In other words the set of constants
$J_{i\,i_3}$ which satisfies  the energy local minimum condition is
not empty and therefore the proposition \textbf{b}) is proved.

So, we have proved the validity of \textbf{a}) and \textbf{b})
propositions. It is obvious that at the simulation of 1$D$ SSC
problem we can by this way fill up 3$D$ space by 1$D$ SSC which is
equivalent to obtaining
 3$D$ spin glass. In case when the number of 1$D$ SSCs is so much
that the directions of spins in 3$D$ space are distributed
isotropically and homogeneous, the statistical properties of both
problems (3$D$ spin glass and 1$D$ SSCs nonideal ensemble) will be
obviously identic.

 The theorem is proved.

\section{Results of Parallel Simulations}
\begin{figure}\center
\includegraphics[height=90mm,width=90mm]{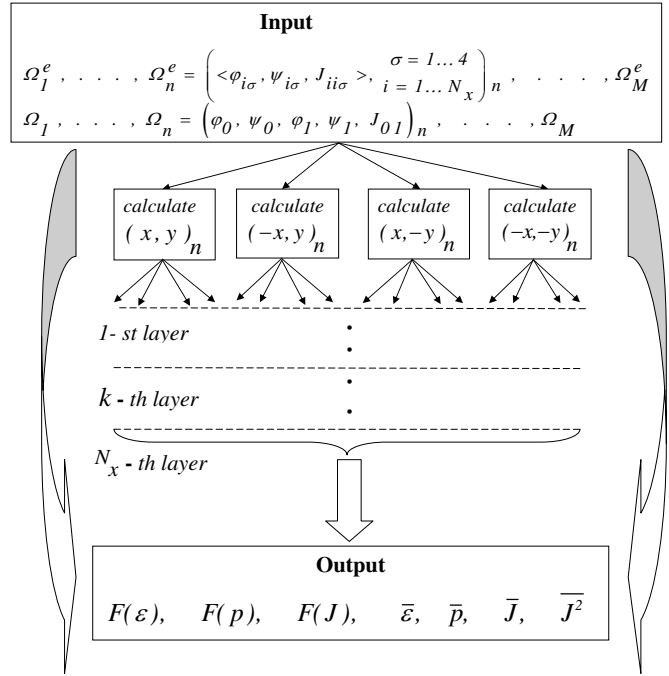}
 \caption{The algorithm of parallel simulation of statistical
parameters of disordered $1D$ SSCs nonideal ensemble. The symbol
$\Omega_n^e$ describes the input  of environment, $M$ is a number of
simulation or overall number of spin-chains in the nonideal
ensemble, $N_x$ is a number of spins in chain.
 }
 \end{figure}

\begin{figure}\center
\includegraphics[height=70mm,width=150mm]{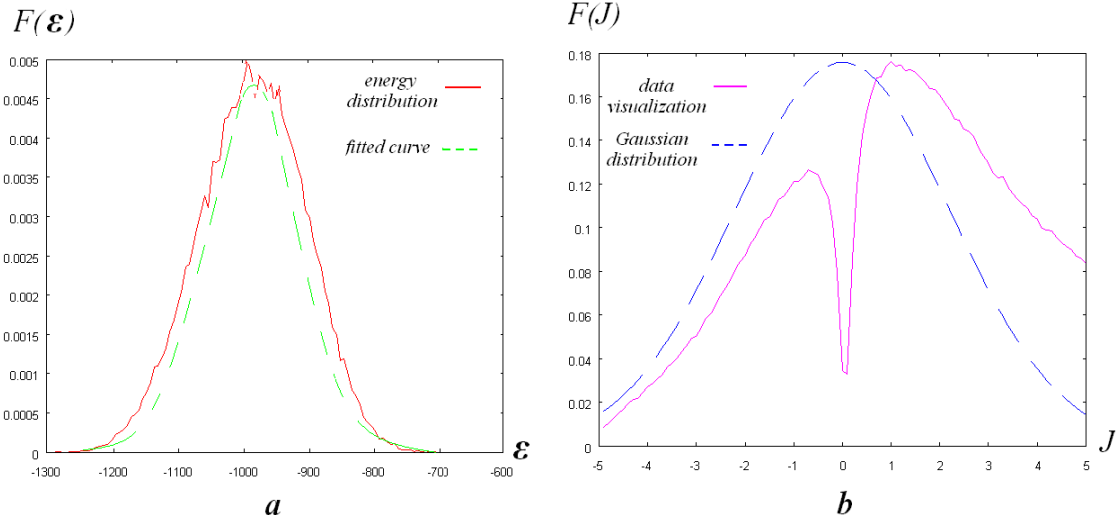}
\caption{On figure \emph{a}) the energy distribution of $1D$
nonideal ensemble of SSCs with $10^3$ length is shown. The red line
shows a numerical data visualization while the green one illustrates
its fitting by Gaussian function. On figure \emph{b}) the
visualization of numerical data of spin-spin interaction constants
(pink line) is shown while the blue one denotes Gaussian
distribution. The analysis of the numerical data proves, that the
green curve is not analytic function and by the character is the
L\'{e}v's skew $\alpha$-stabile distribution function.}
 \end{figure}
One important consequence of the theorem is that for the numerical
simulation of the problem we can use the algorithm for solving the
direct problem. Obviously, a large number of independent
computations of 1$D$ SSC which can be carried out in parallel and in
statistical sense make it equivalent to the problem of 3$D$
spin-glass. This approach considerably reduces the amount of needed
computations and helps us effortlessly simulate  statistical
parameters of 3$D$ spin glasses of large size.

 The strategy of simulation consists of the
following steps (see Fig. 4). At first, the angular configurations
of four spin-chains are randomly generated which form random
environment of the spin-chain which we plan to construct later. On a
following step a set of random constants $J_{i\,i_\sigma}$ are
generated, which characterizes the interactions between the random
environment and the spin-chain. The interaction constants are
generated by Log-normal distribution. The angular configurations of
the random environment are generated the same way as it is described
in \cite{GAS}. Now when the environment and its influence on
disordered 1$D$ SSC are defined, we can go over to the computation
of spin-chain which must satisfy the condition of local energy
minimum. Note that the scheme of further computation of nonideal
ensemble of 1$D$ SSCs (see Fig. 2) is identical to the scheme of the
computation of an ideal ensemble of disordered 1$D$ SSCs (see
\cite{GAS}). Note that all calculations of 1$D$ SSCs nonideal
ensemble are done for spin-chains with $10^3\,d_0$ length which
require huge computational resources.

As the simulations show, for the ensemble which consists of $10^5$
spin-chains, the dimensional effects practically disappear (see
Figs. 5\emph{a}, 5\emph{b} and 6) and the energy distribution
$F(\varepsilon)$ has one global maximum  and is precisely
approximated by Gaussian distribution (see Fig. 5\emph{a}).
\begin{figure}\center
\includegraphics[height=70mm,width=80mm]{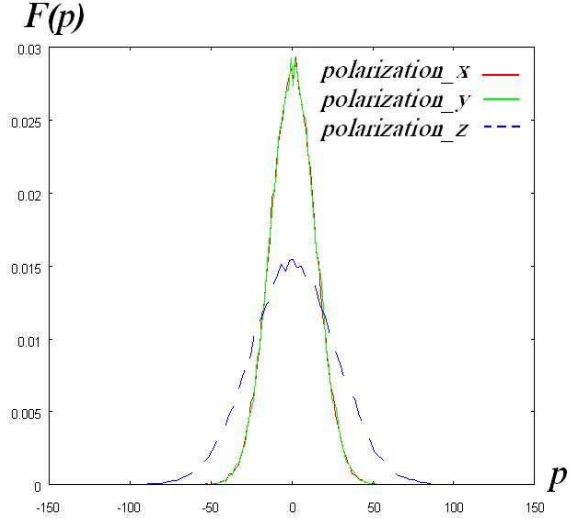}
 \caption{On figure  the  polarization distributions on different
coordinates are shown after $10^5$ simulations.}
\end{figure}

Mean values of  polarizations on coordinates are  not very  small,
especially when it comes to coordinate $x$ (thickness of spin glass
layer): $p_x=-0.13508, \,p_y=0.036586,\, p_z=-0.059995$ and
correspondingly the average energy of $3D$ SSC is equal to
$\bar{\varepsilon}=-990.88$, where $\bar{p}=\int_{-\infty}^{+\infty}
F(p)pdp$, $p=(p_x,\,p_y,\,p_z)$, $\bar{\varepsilon}=\int_{-\infty}^0
F(\varepsilon)\varepsilon d\varepsilon$ and $F$ is the distribution
function. As our numerical investigations have shown on the example
of systems where thickness of spin glass layer is not so large
$\propto $ $25d_0 \div 100d_0$, for a full self-averaging of
superspin it is necessary to make $\propto N_x^2$ simulations. In
other words,  the system can be fully ergodic in considered case if
we continue the numerical simulations of the spin-chains up to
$\propto 10^6$ times.

 It is analytically proved and also the parallel
simulation results show that the spin-spin interaction constant
cannot be described by Gauss-Edwards-Anderson distribution (see Fig.
5\emph{b}). It essentially differs from the normal Gaussian
distribution model and can be approximated precisely by L\'{e}vy
skew alpha-stable distribution function.
 Let us recall that L\'{e}vy
skew alpha-stable distribution is a continuous probability and a
limit of certain random process $X(\alpha,\beta,\gamma,\delta;k)$,
where the parameters correspondingly describe an index of stability
or characteristic exponent $ \alpha\in (0;2]$, a skewness parameter
$\beta\in [-1;1]$, a scale parameter $\gamma > 0$, a location
parameter $\delta \in \textbf{R}$ and an integer $k$ which shows the
certain parametrization (see \cite{Ibragimov,Nolan}). Let us note,
that the mean of distribution and its variance are infinite.
However, taking into account that spin-spin interaction constant has
limited value in real physical systems, it is possible to calculate
distribution mean and its variance. In particular if $J\in [-5,+5]$
then $\overline{J}=0.89717$ and $\overline{J^2}=5.3382$.

In the work are also presented  polarization distributions on
different coordinates (see Fig. 6).  As for the polarization
distributions, they are obviously very symmetric by coordinates in
the considered case (see Fig. 6).

One of the advantages of the developed algorithm is that we are able
to take into account the branching solutions  at the successive
constructing of the spin-chain (see Fig. 7). As calculations show,
the number of branching solutions $\nu$ for spin-chains of length
$10^3\,d_0$  is not more than 25.  At the simulation process only
those spin-chains are considered for which Silvester conditions are
satisfied on each node. If on some node the conditions are not
satisfied we try to regenerate $J_{i\,i+1}$ in order to obtain a new
solution. However, if the solution is not found after large quantity
simulations it means that the weight of these solutions all  are
extremely small and  further simulations of these spin-chains are
unpractical.
\begin{figure}\center
\includegraphics[height=80mm,width=150mm]{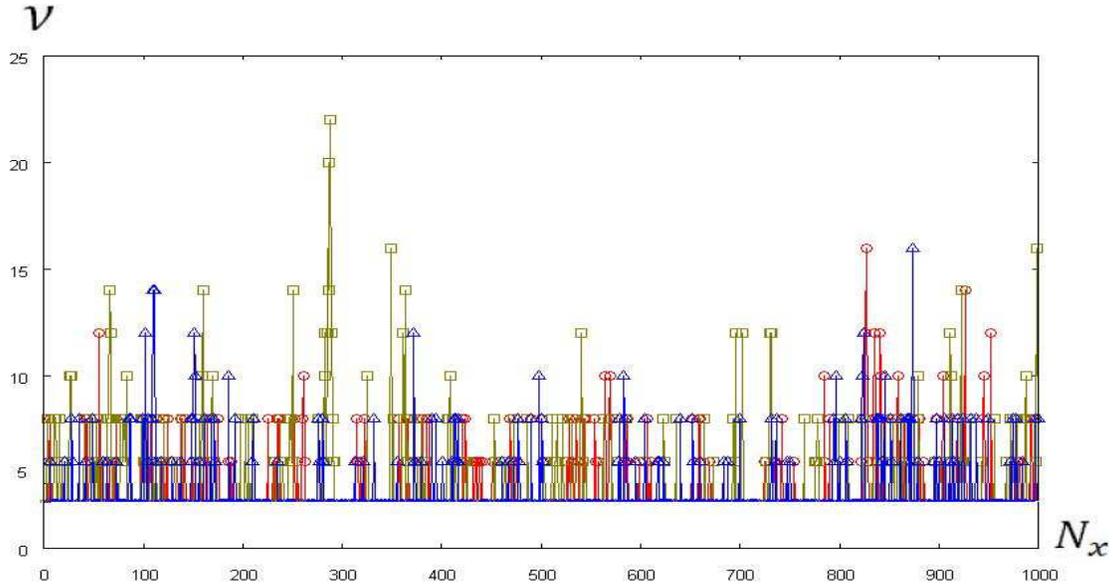}
 \caption{ On figure  the
number of branching  of solutions $\nu$ is shown along with the
spin-chain length depending on different initial conditions which
are indicated by way of various colors. }
\end{figure}
Thus, in case when the ensemble consists of a large number of
spin-chains, the self-averaging of superspin (sum vector of
spin-chain) in 3$D$ space occurs with high accuracy. It is important
to note, that the summation procedure on the number of spins in
chain or on  the number of spin-chains in ensemble, is similar to
the procedure of averaging by the natural parameter or "timing" in
the dynamical system. The latter means that at defined space scales
of spin glasses it is possible to introduce the concept of
ergodicity for both separate spin-chains and ensemble as a whole.

\section{Partition function}

The main object of investigation of statistical mechanics,
information science, probability theory and etc., is the partition
function which is defined for classical many particle case in
configuration space as follows \cite{Wannier}:
\begin{equation}
Z(\beta)=\int \exp\bigl[-\beta
H(\{\textbf{r}\})\bigr]d\textbf{r}_1d\textbf{r}_2...,\qquad
\beta=\frac{1}{k_BT}, \label{26}
\end{equation}
where $k_B$ is the Boltzmann constant and $T$ is the thermodynamic
temperature. Obviously, when the number of spins or spin-chains  in
the system are large we  can  consider integral (\ref{26})  as a
functional integral. In any case the number of integration in
expression (\ref{26}) as a rule is very large for many tasks  and
the main problem lies in the correct calculation of this integral.
However, in the representation of (\ref{26}) configurations of
spin-chains  that are not physically realizable obviously make a
contribution. Moreover, the weight of these configurations is not
known in general scenario and it is unclear how to define it. With
this in mind and also taking into account the ergodicity of the spin
glass in the above mentioned sense, we can define the partition
function as:
\begin{equation}
Z_\ast(\beta;N_x)=\int_{-\infty}^{\,0} \exp\bigl[\beta
\varepsilon\bigr]F(\varepsilon;N_x)d\varepsilon, \label{27}
\end{equation}
where $F(\varepsilon;N_x)$ is the energy distribution function in
nonideal ensemble of 1$D$ SSCs  with certain length $N_x$ (see also
definition (\ref{06})).

Now we can define the Helmholtz free energy for ensemble of 1$D$
SSCs by two different ways. Using standard definition for Helmholtz
free energy we can write:
\begin{equation}
Q(\beta; N_x)=-\frac{1}{N_x\beta}\ln\bigl[Z(\beta; N_x)\bigr], \quad
Q_\ast(\beta; N_x)=-\frac{1}{N_x\beta}\ln\bigl[Z_\ast(\beta;
N_x)\bigr]. \label{28}
\end{equation}
Note that the dependence on $N_x$ of the expressions in (\ref{28})
arises due to  the finite  layer width. In particular, using the
expression of partition function (\ref{26}) we can  find the average
value of free energy coming on one spin in the chain (see also
\cite{Ziman}):
\begin{equation}
Q(\beta;N_x)=-\frac{1}{N_x\beta}\biggl\langle\sum_{i=0}^{N_x-1}\ln\biggl[\frac{\sinh
x_i}{x_i}\biggr]\biggr\rangle,\qquad x_i=J_{i\,i+1}\beta, \label{29}
\end{equation}
where $\bigl\langle\,...\,\bigr\rangle$ designates averaging by 1$D$
SSCs ensemble. Now the main problem is the investigation of behavior
of free energy subject to the parameter $\beta$. Correspondingly we
can define two forms of free energy derivatives:
$$
q(\beta;N_x)=\frac{\partial Q(\beta; N_x)}{\partial
\beta}=\frac{1}{N_x\beta^2}\,\biggl\langle\sum_{i=0}^{N_x-1}\biggl\{1+
\ln\biggl[\frac{\sinh{x_i}}{x_i}\biggr]-x_i\coth{x_i}\biggr\}\biggr\rangle,
$$
\begin{figure}\center
\includegraphics[height=70mm,width=80mm]{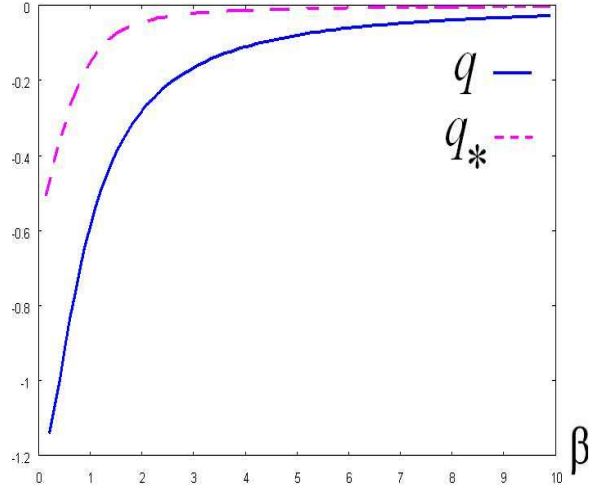}
 \caption{ On the figure two derivatives of Helmholtz's free energy
are shown depending on reverse temperature $ \ beta $ which have
been calculated by formulas (\ref{30}) after $ 10 ^ 6 $ simulations.
The figure shows that at low temperatures the curves approach each
other which is rather natural, resulting from the decrease of the
system's entropy.}
\end{figure}
\begin{eqnarray}
 q_\ast(\beta;N_x)=\frac{\partial Q_\ast(\beta; N_x)}{\partial
\beta}=\frac{1}{N_x\beta}\biggl\{\frac{1}{\beta}\ln
Z_\ast(\beta;N_x)-\frac{Z_{\ast;\,\beta}(\beta;N_x)}{Z_\ast(\beta;N_x)}\biggr\},
\label{30}
\end{eqnarray}
where $Z_{\ast;\,\beta}(\beta;N_x)=\partial
Z_{\ast}(\beta;N_x)/\partial \beta$.

\section{Conclusion}
A new parallel algorithm is developed for the simulation of the
classical 3$D$ spin glasses. It is shown that 3$D$ spin glasses can
be investigated by the help of an auxiliary Heisenberg Hamiltonian
(\ref{01}). The system of recurrent transcendental equations
(\ref{03}) and Silvester conditions (\ref{04}) are obtained  by
using this Hamiltonian. Let us note that exactly similar  equations
of stationary points (\ref{03}) also can be obtained if the full
3$D$ Hamiltonian (see the first unnumbered formula) is used in the
framework of short-range interaction model. That allows us step by
step construct spin-chain of the specified length with taking into
account the random surroundings. It is proved that at the limit of
Birkhoff's ergodic hypothesis performance, 3$D$ spin glass can be
generated by Hamiltonian of disordered 1$D$ SSC with random
environment. We have proved that it is always possible to construct
spin-chain in any given random environment which will be in ground
state energy (direct problem). We have also proved the inverse
problem, namely, every spin-chain of the random environment can be
surrounded by an environment so that it will be the solution in the
ground state.  In the work all the necessary numerical data were
obtained by way of large number of parallel simulations of the
auxiliary problem in order to construct all the statistical
parameters of 3$D$ spin glass at the limit of ergodicity of 1$D$
SSCs nonideal ensemble. As numerical simulations show, the
distributions of all statistical parameters become stable after
$\propto N_x^{\,2}$ independent calculations which are realized in
parallel. The idea of 1$D$ spin-chains parallel simulations, based
on this simple and clear logic, greatly simplifies the calculations
of 3$D$ spin glasses which are still considered as a subset of
difficult simulation problems. Let us note that computation of
spin-spin interactions distribution function from the first
principals of the classical mechanics  is very important result of
this work. As analysis show, the distribution is not an analytic
function. It is from the class of L\'{e}vy functions which does not
have variance $\overline{J^{2}}$ and mean value $\bar{J}$.

Despite the absence of calculations by other methods, it is obvious
that the developed scheme of calculations should differ from other
algorithms, including the algorithms which are based on Monte Carlo
simulation method \cite{Metropolis},  by the accuracy and
efficiency. We were once again convinced in the accuracy and
efficiency of the algorithm after analyzing the results of different
numerical experiments by modeling the statistical parameters of 3$D$
spin-glass system which are presented in figures 5 a, b and 6.

In the work a new way of partition function construction
(configuration integral) is proposed in the form of one-dimensional
integral of the energy distribution, which unlike the usual
definitions does not include physically unrealizable spin-chains
configurations (see difference of free energy derivatives on Fig.
8). It is obvious that the new definition of partition function is
more correct and in addition it is very simple for computation.

Finally,  the developed method can be generalized for the cases of
external fields which will allow us investigate a large number of
dynamical problems including critical properties of 3$D$ classical
spin glasses.

\section{Acknowledgment} The work is partially supported by RFBR grants 10-01-00467-a, 11-01-0027-a.

\end{document}